\documentclass[reprint,noshowpacs,noshowkeys,prd,balancelastpage,nofootinbib]{revtex4}
\usepackage[utf8]{inputenc}
\usepackage{color}
\usepackage{float}
\usepackage{amsmath}
\usepackage{amssymb}
\usepackage{graphicx}
\usepackage[unicode=true,
 bookmarks=false,
 breaklinks=false,pdfborder={0 0 1},backref=section,colorlinks=true]
 {hyperref}
\hypersetup{
 linkcolor=blue,urlcolor=blue,citecolor=red}

\makeatletter
\@ifundefined{textcolor}{}
{%
 \definecolor{BLACK}{gray}{0}
 \definecolor{WHITE}{gray}{1}
 \definecolor{RED}{rgb}{1,0,0}
 \definecolor{GREEN}{rgb}{0,1,0}
 \definecolor{BLUE}{rgb}{0,0,1}
 \definecolor{CYAN}{cmyk}{1,0,0,0}
 \definecolor{MAGENTA}{cmyk}{0,1,0,0}
 \definecolor{YELLOW}{cmyk}{0,0,1,0}
}

\usepackage{amsfonts}
\usepackage{mdframed}
\usepackage{footnote}
\usepackage[font=footnotesize,it]{caption}
\usepackage{tikz}
\usepackage{tikz-3dplot}

\makeatother

\begin{document}
\title{Black Holes in Double-Logarithmic Nonlinear Electrodynamics}
\author{Ibrahim Gullu}
\email{ibrahim.gullu@emu.edu.tr}

\author{S. Habib Mazharimousavi}
\email{habib.mazhari@emu.edu.tr}

\affiliation{Department of Physics, Faculty of Arts and Sciences, Eastern Mediterranean
University, Famagusta, North Cyprus via Mersin 10, Turkey}
\date{\today }
\begin{abstract}
The electric and magnetic black hole solutions are found by coupling
the recently introduced nonlinear electrodynamics (NED) model, called
``double logharitmic nonlinear electrodynamics'' with cosmological
Einstein gravity. The solutions become Reissner-Nordstrom (RN) black
hole in the weak field limit and asymptotically. The electric solution
is expressed as an integral equation while the magnetic black hole
solution is expressed in terms of elementary functions. Hence, the
thermodynamic structure of the magnetic black hole solution is analyzed
by deriving the modified Smarr's formula and studying the first law
of thermodynamics. Moreover, its stability is investigated by deriving
the heat capacity. 
\end{abstract}
\maketitle

\section{Introduction}

In early time cosmology the problem of big bang singularity and inflation
are still contemporary. Nonlinear electrodynamics (NED) maybe a solution
to these problems. The gravitational fields and electromagnetic fields
were very strong in creation of the universe. There is an inextricable
relation between strong fields and nonlinearities therefore the nonlinear
effects not only took an important role in early time universe but
also are crucial in understanding black hole singularities.

As a NED Born-Infeld (BI) electrodynamics \cite{Born,Born-Infeld},
whose action comes out from low energy effective action of superstring
theory \cite{Fradkin-Tseytlin,Tseytlin}, solves the problem of singularity
of the electric field at the center of a charged point particle, therefore
the electric energy of charged particles becomes finite. Beside BI,
some models of NED are introduced in \cite{Gitman-Shabad,Kruglov-1,Kruglov-2,Kruglov-3,Gullu-Mazharimousavi}
and these models are free of singularity of the electric field at
the center of charged particles. At the weak field limit BI theory
goes to Maxwell electrodynamics which can be thought as an approximation
of NED.

The properties of NED can be comprehensible in the framework of gravity
since, strong electromagnetic fields are dominant in the early period
of the universe, at the center of black holes and charged particles.
The exact black hole solution in General Relativity (GR) was given
in \cite{Oliveira} by use of a NED Lagrangian which defines a large
class of non-linear theories including BI and Euler-Heisenberg (EH)
electrodynamics whose weak field limit also tends to Maxwell electrodynamics.
Other models of NED coupled to gravity have been studied in \cite{Ayon-Beato-Garcia,Breton-1,Dymnikova,Balart-Vagenas,Kruglov-4}.
The solution of these models are singularity free black hole solutions
which asymptotically behave as the Raisner-Nördstrom (RN) black hole
\cite{Reissner,Nordstrom}. There is already a rich literature in
this respect as contributions of NED models in the formation of different
type black holes \cite{Fernando-Krug}-\cite{Kruglov-13}. Moreover,
instead of dark energy, non-linear electromagnetic fields are introduced
to explain the inflation period in early time universe \cite{Garcia-Breton,Camara-et.al.}.
In addition, some models of NED were used to depict accelerated expansion
of the universe \cite{Elizalde-et.al.,Novello-et.al.-1,Novello-et.al.-2,Vollick,Kruglov-5,Kruglov-6}.
In $\Lambda$-Cold Dark Matter ($\Lambda$CDM) model the expansion
of the universe is driven by the cosmological constant $\Lambda$,
on the other hand the non-zero trace of the energy-momentum tensor
in NED can play the role of cosmological constant \cite{Labun,Schutzhold}.

In this paper we consider a recently introduced NED model \cite{Gullu-Mazharimousavi},
which is constructed by both Maxwell (Lorentz) invariants $\mathcal{F}=F^{\mu\nu}F_{\mu\nu}$
and $\mathcal{G}=F_{\mu\nu}\tilde{F}^{\mu\nu}$ and carries one dimensional
parameter $\beta$, and couple it to the gravitational field. Due
to the technical difficulty of finding a dyonic black hole solution
in which the electric and magnetic fields coexist, we have found electric
and magnetic black hole solutions separately in the absence of the
second invariant term, $\mathcal{G}$, in the model. The solutions
are in the form of modified RN black hole for both cases. We find
the electric type black hole solution by setting the magnetic field
to zero and letting the source to be electrically charged. The solution
is inexact namely, it can only be expressed as an integral expression
therefore its series expansion can be of any use. In the second case
the electric field is set to zero and the source of gravitational
field becomes magnetized and the magnetic black hole solution is exact.
Due to this property, the magnetic black hole solution is analyzed
in more detail. The first law of thermodynamics, Hawking radiation,
the extension of Smarr's formula and thermal stability are worked
out for the magnetic black hole solution \cite{Smarr,Breton-2,Rasheed,Zhang-Gao,Hu-Kuang-Ong,Wang-Wu-Yang}.

The layout of the paper is as follows: In Section 2 we review the
NED model under consideration. In Section 3 electric and magnetic
type solutions are found by coupling minimally nonlinear electromagnetic
fields to gravitational fields. The thermodynamic properties of the
magnetic type solution is investigated in Section 4. Thermal stability
of the magnetic black hole constitute the subject for Section 5. The
paper is brought to completion with the Conclusion in Section 6.

In the continuation, we take $\hbar=c=\varepsilon_{0}=\mu_{0}=1$
and the metric with mostly plus signature. The coordinates are defined
as $x^{\mu}=\left(t,r,\theta,\phi\right)$. Greek indices run from
$0$ to $3$ and Latin indices run from $1$ to $3$.

\section{The New NED Model:}

The Double-Logarithmic Lagrangian \cite{Gullu-Mazharimousavi} is
given by
\begin{align}
\mathcal{L}_{e} & =\frac{1}{2\beta}\left[\left(1-\sqrt{-2\beta\mathcal{F}}\right)\ln\left(1-\sqrt{-2\beta\mathcal{F}}\right)+\left(1+\sqrt{-2\beta\mathcal{F}}\right)\ln\left(1+\sqrt{-2\beta\mathcal{F}}\right)\right],\label{Lagrangian}
\end{align}
in which $\mathcal{F}=F_{\mu\nu}F^{\mu\nu}=2\left(\boldsymbol{B}^{2}-\boldsymbol{E}^{2}\right)$
is the Maxwell invariant and the electromagnetic strength tensor is
given in terms of the gauge potential $F_{\mu\nu}=\partial_{\mu}A_{\nu}-\partial_{\nu}A_{\mu}$
which in the matrix form reads 
\begin{align}
F_{\mu\nu}= & \left(\begin{array}{cccc}
0 & E_{r} & 0 & 0\\
-E_{r} & 0 & 0 & 0\\
0 & 0 & 0 & -B_{r}r^{2}\sin\theta\\
0 & 0 & B_{r}r^{2}\sin\theta & 0
\end{array}\right),\label{electromagnetic_tensor-1}\\
F^{\mu\nu}= & \left(\begin{array}{cccc}
0 & -E_{r} & 0 & 0\\
E_{r} & 0 & 0 & 0\\
0 & 0 & 0 & -\frac{B_{r}}{r^{2}\sin\theta}\\
0 & 0 & \frac{B_{r}}{r^{2}\sin\theta} & 0
\end{array}\right),\label{electromagnetic_tensor-2}
\end{align}
where the electric field and magnetic field are static and depend
only on the radial coordinate. In the following computation we need
also, 
\begin{align}
F_{\mu\lambda}F^{\nu\lambda}= & \left(\begin{array}{cccc}
-E_{r}^{2} & 0 & 0 & 0\\
0 & -E_{r}^{2} & 0 & 0\\
0 & 0 & B_{r}^{2} & 0\\
0 & 0 & 0 & B_{r}^{2}
\end{array}\right).\label{identity_of_emst}
\end{align}
In the weak field limit $\beta\rightarrow0$, (\ref{Lagrangian})
reduces to the linear Maxwell Lagrangian 
\begin{equation}
-\mathcal{F}+\frac{1}{3}\mathcal{F}^{2}\beta-\frac{4}{15}\mathcal{F}^{3}\beta^{2}+O[\beta]^{5/2}.\label{b_zero_limit_s_0}
\end{equation}
where the zeroth order term is the usual Maxwell's theory which can
be achieved when $\beta=0$. The first order correction to the Maxwell's
theory is $\frac{1}{3}\mathcal{F}^{2}$. The Lagrangian for the gravity
part is the cosmological Einstein-Hilbert Lagrangian 
\begin{equation}
\mathcal{L}_{g}=R-2\Lambda,\label{Lagrangian_G}
\end{equation}
where $\Lambda$ is the cosmological constant and the action in which
the electromagnetic field couples minimally to the gravity field reads
as 
\begin{equation}
I=\int d^{4}x\sqrt{-g}\left(\frac{1}{2\kappa}\mathcal{L}_{g}+\mathcal{L}_{e}\right),\label{Action}
\end{equation}
where $\kappa=8\pi G$ and $G$ is the four dimensional Newton's constant.
The static spherically symmetric line element is chosen to be 
\begin{align}
ds^{2}= & -f\left(r\right)dt^{2}+\frac{1}{f\left(r\right)}dr^{2}+r^{2}d\Omega^{2},\label{spacetime_metric}
\end{align}
in which $d\Omega=d\theta^{2}+\sin^{2}\theta d\phi^{2}$. The field
equations of gravitational and electromagnetic fields are 
\begin{align}
\mathcal{G}_{\mu}^{\nu}= & \kappa T_{\mu}^{\nu},\label{eom_metric}
\end{align}
where the cosmological Einstein tensor reads $\mathcal{G}_{\mu}^{\nu}=G_{\mu}^{\nu}-\Lambda\delta_{\mu}^{\nu}$
with the Einstein tensor $G_{\mu}^{\nu}=R_{\mu}^{\nu}-\frac{1}{2}\delta_{\mu}^{\nu}R$
and the energy-momentum tensor is defined to be
\begin{equation}
T_{\mu}^{\nu}\equiv\mathcal{L}\delta_{\mu}^{\nu}-4\mathcal{L}_{\mathcal{F}}F_{\mu\lambda}F^{\nu\lambda}\label{eq:EMT}
\end{equation}
which results in

\begin{align}
T_{\mu}^{\nu}= & \frac{1}{2\beta}\left[\left(1-\sqrt{-2\beta\mathcal{F}}\right)\ln\left(1-\sqrt{-2\beta\mathcal{F}}\right)+\left(1+\sqrt{-2\beta\mathcal{F}}\right)\ln\left(1+\sqrt{-2\beta\mathcal{F}}\right)\right]\delta_{\mu}^{\nu}\nonumber \\
 & -\frac{2F_{\mu\lambda}F^{\nu\lambda}}{\sqrt{-2\beta\mathcal{F}}}\ln\left(\frac{1-\sqrt{-2\beta\mathcal{F}}}{1+\sqrt{-2\beta\mathcal{F}}}\right),\label{energy_momentum}
\end{align}
and the second field equation reads

\begin{align}
\partial_{\alpha}\left\{ \frac{\sqrt{-g}}{\sqrt{-2\beta\mathcal{F}}}\ln\left(\frac{1-\sqrt{-2\beta\mathcal{F}}}{1+\sqrt{-2\beta\mathcal{F}}}\right)F^{\alpha\beta}\right\} = & 0\label{eom_gauge_potential}
\end{align}
after varying the action (\ref{Action}) with respect to the gauge
potential $A_{\mu}$. For the metric (\ref{spacetime_metric}) the
components of the cosmological Einstein tensor becomes 
\begin{align}
\mathcal{G}_{0}^{0}=\mathcal{G}_{1}^{1}= & \frac{rf'\left(r\right)+f\left(r\right)-1-r^{2}\Lambda}{r^{2}}\nonumber \\
\mathcal{G}_{2}^{2}=\mathcal{G}_{3}^{3}= & \frac{rf''\left(r\right)+2f'\left(r\right)-2\Lambda r}{2r}\label{Einstein_tensor}
\end{align}
which will be needed for later use.

\section{Electric Part:}

In this section we find the electric field of a point charge and the
electric black hole solution for $B_{r}=0$ which means we have only
electric monopole sitting at the origin.

\subsection{The Electric Field:}

The electric field of a point charge can be found by use of (\ref{eom_gauge_potential}).
The Maxwell invariant depends only on the Electric field, $\mathcal{F}=-2E_{r}^{2}$,
and the electric field has just radial component $\boldsymbol{E}=\left(E_{r}\left(r\right),\,0,\,0\right)$.
The determinant of the metric tensor (\ref{spacetime_metric}) is
\begin{align}
\sqrt{-g}= & r^{2}\sin\theta.\label{metric_det}
\end{align}
Once the index $\beta$ is taken to be zero, then (\ref{eom_gauge_potential})
reduces to

\begin{align}
\partial_{r}\left\{ \frac{r^{2}}{\sqrt{\beta}}\left(\tanh^{-1}\left(\sqrt{\beta}E_{r}\left(r\right)\right)\right)\right\}  & =0.\label{zeroth_diff_eq}
\end{align}
The terms inside the parenthesis in (\ref{zeroth_diff_eq}) are equal
to a constant. Then, the electric field of a point particle reads
\begin{align}
E_{r}\left(r\right)= & \frac{1}{2\sqrt{\beta}}\tanh\left(\frac{2\sqrt{\beta}q}{r^{2}}\right),\label{electric_field}
\end{align}
where $q$ is a constant corresponding to the electric charge of the
black hole. In the weak field limit $\beta\rightarrow0$ the electric
field reduces to Coulomb's field i.e., $E_{r}=\frac{q}{r^{2}}$.

\subsection{The Electric Black Hole Solution:}

In this part we will solve the Einstein field equations (\ref{eom_metric})
upon use of (\ref{spacetime_metric}). The energy-momentum tensor
reads explicitly 
\begin{align}
T_{\mu}^{\nu}= & \mathcal{A}\delta_{\mu}^{\nu}-\mathcal{B}F_{\mu\lambda}F^{\nu\lambda},\label{T_mn_1}
\end{align}
where we have defined 
\begin{align}
\mathcal{A}\equiv & \frac{1}{2\beta}\left[\left(1-2\sqrt{\beta}E_{r}\right)\ln\left(1-2\sqrt{\beta}E_{r}\right)+\left(1+2\sqrt{\beta}E_{r}\right)\ln\left(1+2\sqrt{\beta}E_{r}\right)\right],\label{A}\\
\mathcal{B}\equiv & \frac{1}{\sqrt{\beta}E_{r}}\ln\left(\frac{1-2\sqrt{\beta}E_{r}}{1+2\sqrt{\beta}E_{r}}\right).\label{B}
\end{align}
Then, (\ref{eom_metric}) reduces to the first order differential
equation given by
\begin{align}
f^{\prime}\left(r\right)r+f\left(r\right)-\Lambda r^{2}-1= & \frac{\kappa r^{2}}{2\beta}\ln\left(1-4\beta E_{r}^{2}\right).\label{first_diff_eq}
\end{align}
Rewriting (\ref{first_diff_eq}) as 
\begin{align}
\left(f\left(r\right)r\right)^{\prime}= & \frac{\kappa r^{2}}{2\beta}\ln\left(\text{sech}^{2}\left(\frac{2\sqrt{\beta}q}{r^{2}}\right)\right)+\Lambda r^{2}+1\label{first_diff_eq_modified}
\end{align}
where we have used (\ref{electric_field}) and the trigonometric identity
$\text{sech}^{2}x=1-\tanh^{2}x$. Integrating both sides of (\ref{first_diff_eq_modified})
one gets the solution of electric black hole as follows 
\begin{align}
f\left(r\right)= & \frac{\kappa}{2\beta r}\int r^{2}\ln\left(\text{sech}^{2}\left(\frac{2\sqrt{\beta}q}{r^{2}}\right)\right)dr+\frac{\Lambda r^{2}}{3}+1-\frac{2GM}{r},\label{first_diff_eq_integrating}
\end{align}
where the integration constant is defined as 
\begin{equation}
C=2GM,\label{integration_constant}
\end{equation}
in which $M$ represents the Schwarzschield mass when there is no
electromagnetic field. This solution can be numerically analyzed because
the integral can not be expressed in terms of elementary functions.
After expanding (\ref{first_diff_eq_integrating}) in Taylor series
expansion around $\beta=0$ up to the first order we get 
\begin{align}
f\left(r\right)= & 1-\frac{2MG}{r}+\frac{2\kappa q^{2}}{r^{2}}+\frac{\Lambda}{3}r^{2}-\frac{4\kappa q^{4}\beta}{15r^{6}}-\frac{2G}{\beta r}\int r^{2}\mathcal{O}\left(\beta\right)^{\frac{3}{2}}dr.\label{solution_2}
\end{align}
Setting $\beta=0$ the solution (\ref{solution_2}) reduces to 
\begin{align}
f\left(r\right)= & 1-\frac{2MG}{r}+\frac{2\kappa q^{2}}{r^{2}}+\frac{\Lambda}{3}r^{2},\label{solution_3}
\end{align}
which is the RN solution not only in the weak field limit but also
asymptotically. Since, this solution can not be written in terms of
elementary functions we have to confine ourselves to these limits.
In the next section, we shall find an exact magnetic black hole solution
whose physical properties can be analyzed.

\section{Magnetic Part:}

The Bianchi identity for $E_{r}=0$ implies that 
\[
B_{r}=\frac{P}{r^{2}}
\]
in which $P$ is the magnetic monopole charge. The equation of motion
(\ref{eom_metric}) with $\mathcal{F}=\frac{2P^{2}}{r^{4}}$ reads
\begin{align}
\frac{1}{\kappa}\mathcal{G}_{\mu}^{\nu}= & \mathcal{C}\delta_{\mu}^{\nu}-\mathcal{D}F_{\mu\lambda}F^{\nu\lambda}\label{eom_magnetic}
\end{align}
where the right hand side corresponds to the energy-momentum tensor
with the following defined parameters 
\begin{align}
\mathcal{C}\equiv & \frac{1}{2\beta}\left[\left(1-\frac{2\sqrt{-\beta}P}{r^{2}}\right)\ln\left(1-\frac{2\sqrt{-\beta}P}{r^{2}}\right)+\left(1+\frac{2\sqrt{-\beta}P}{r^{2}}\right)\ln\left(1+\frac{2\sqrt{-\beta}P}{r^{2}}\right)\right]\nonumber \\
\mathcal{D}\equiv & \frac{1}{\sqrt{-\beta}\frac{P}{r^{2}}}\ln\left(\frac{1-2\sqrt{-\beta}\frac{P}{r^{2}}}{1+2\sqrt{-\beta}\frac{P}{r^{2}}}\right).\label{definition_of_C_D}
\end{align}
The differential equation, we get from (\ref{eom_magnetic}) is

\begin{align}
f^{\prime}\left(r\right)r+f\left(r\right)= & \frac{\kappa r^{2}}{\beta}\left(\ln\sqrt{1+\frac{4\beta P^{2}}{r^{4}}}-2\frac{\sqrt{\beta}P}{r^{2}}\arctan\left(2\frac{\sqrt{\beta}P}{r^{2}}\right)\right)+1+\Lambda r^{2}\label{second_diff_eqn}
\end{align}
which admits the solution 
\begin{align}
f\left(r\right)= & -\frac{2MG}{r}+\frac{\Lambda}{3}r^{2}+1-\frac{4\kappa P^{2}}{r_{0}^{2}}\arctan\left(\frac{r_{0}^{2}}{r^{2}}\right)+\frac{2\kappa P^{2}}{3r_{0}^{4}}r^{2}\ln\left(\frac{r^{4}+r_{0}^{4}}{r^{4}}\right)\nonumber \\
 & +\frac{4\sqrt{2}\kappa P^{2}}{3r_{0}r}\left[\arctan\left(1-\frac{\sqrt{2}r}{r_{0}}\right)-\arctan\left(1+\frac{\sqrt{2}r}{r_{0}}\right)\right]\nonumber \\
 & +\frac{2\sqrt{2}\kappa P^{2}}{3r_{0}r}\ln\left(\frac{r^{2}+\sqrt{2}r_{0}r+r_{0}^{2}}{r^{2}-\sqrt{2}r_{0}r+r_{0}^{2}}\right),\label{exact_solution_magnetic_part}
\end{align}
in which $r_{0}^{2}=2P\sqrt{\beta}$ and we have defined the integration
constant as (\ref{integration_constant}) where $M$ refers to the
total mass of the black hole. Once, the weak field limit is taken
i.e., $\beta\rightarrow0$, after expanding (\ref{exact_solution_magnetic_part})
around $\beta=0$ we get the solution for the magnetic black hole
as 
\begin{align}
f\left(r\right)= & 1-\frac{2GM}{r}+\frac{2\kappa P^{2}}{r^{2}}+\frac{r^{2}\Lambda}{3},\label{magnetic_bh}
\end{align}
which is the RN solution with the ADM mass $M$. Asymptotically, when
$r\rightarrow\infty$, the magnetic black hole solution reads 
\begin{align}
f\left(r\right)= & 1+\frac{\Lambda}{3}r^{2}-2G\left(M+\frac{2\kappa\sqrt{2}P^{2}\pi}{3Gr_{0}}\right)\frac{1}{r}+\frac{2\kappa P^{2}}{r^{2}}+\mathcal{O}\left(r^{3}\right),\label{Assymptotic_magnetic_bh_solution}
\end{align}
where $M_{\text{ADM}}$, the apparent mass, becomes 
\begin{align}
M_{\text{ADM}}= & M+\frac{2\kappa\sqrt{2}\pi P^{2}}{3r_{0}G},\label{Total_mass}
\end{align}
such that the second term in the right hand side is the share of NED's
field mass. Since, the last term in (\ref{Assymptotic_magnetic_bh_solution})
goes to zero at $r\rightarrow\infty$ the solution reduces to (anti)-de
Sitter-Schwarzschild black hole and without the cosmological constant
it reduces to Schwarzschild black hole as expected. Around the origin
$r\rightarrow0$ the solution (\ref{exact_solution_magnetic_part})
behaves as 
\begin{align}
f\left(r\right)= & 1-\frac{\pi\kappa P}{\sqrt{\beta}}-\frac{2GM}{r}+\mathcal{O}\left(r^{3}\right),\label{Magnetic_BH_around_origin}
\end{align}
which is apparently singular due to the constant term $-\frac{\pi\kappa P}{\sqrt{\beta}}$,
a global monopole type constant which gives rise to a deflection angle
\cite{Barriola-Vilenkin}.

\section{The First Law of Thermodynamics and Smarr's Formula:}

In this section the thermodynamic structure of (\ref{exact_solution_magnetic_part})
is analyzed. For this reason, the Smarr's formula is constructed by
use of the Euler's homogeneous function theorem stating that if $f\left(\lambda^{i}x,\lambda^{j}y,\lambda^{k}z\right)=\lambda^{l}f\left(x,y,z\right)$,
where $\lambda$ is a constant and $\left(i,j,k,l\right)$ are integer
powers, then $lf\left(x,y,z\right)=ix\left(\frac{\partial f}{\partial x}\right)+jy\left(\frac{\partial f}{\partial y}\right)+kz\left(\frac{\partial f}{\partial z}\right)$.
First let us find the mass of the black hole using (\ref{exact_solution_magnetic_part})
which becomes zero at the horizon $r=r_{h}$ 
\begin{align}
M= & \frac{\sqrt{2}P^{3/2}\pi}{3\beta^{1/4}}+\frac{\Lambda}{6G}r_{h}^{3}+\frac{r_{h}}{2G}-\frac{P\beta^{\frac{1}{2}}r_{h}}{\beta}\arctan\left(\frac{P\beta^{\frac{1}{2}}}{r_{h}^{2}}\right)+\frac{r_{h}^{3}}{6\beta}\ln\left(\frac{r_{h}^{4}+P^{2}\beta}{r_{h}^{4}}\right)\nonumber \\
 & +\frac{\sqrt{2}P^{\frac{3}{2}}\beta^{\frac{3}{4}}}{3\beta}\arctan\left(1-\frac{\sqrt{2}r_{h}}{P^{\frac{1}{2}}\beta^{\frac{1}{4}}}\right)-\frac{\sqrt{2}P^{\frac{3}{2}}\beta^{\frac{3}{4}}}{3\beta}\arctan\left(1+\frac{\sqrt{2}r_{h}}{P^{\frac{1}{2}}\beta^{\frac{1}{4}}}\right)\nonumber \\
 & +\frac{\sqrt{2}P^{\frac{3}{2}}\beta^{\frac{3}{4}}}{6\beta}\ln\left(\frac{P\beta^{\frac{1}{2}}+\sqrt{2}P^{\frac{1}{2}}\beta^{\frac{1}{4}}r_{h}+r_{h}^{2}}{P\beta^{\frac{1}{2}}-\sqrt{2}P^{\frac{1}{2}}\beta^{\frac{1}{4}}r_{h}+r_{h}^{2}}\right),\label{mass_of_bh}
\end{align}
where $r_{h}$ is the radius of the horizon. The Hawking-Bekenstein
entropy is $S=\frac{\pi r_{h}^{2}}{G}$, then $r_{h}=\sqrt{\frac{SG}{\pi}}$.
The ADM mass of the black hole can be rewritten in terms of the entropy,
magnetic charge and the $\beta$ parameter as 
\begin{align}
M\left(S,P,\beta,\Lambda\right)= & \frac{2P^{2}\kappa\pi}{3\sqrt{P\sqrt{\beta}}G}+\frac{S\Lambda}{6\pi}\sqrt{\frac{SG}{\pi}}+\frac{1}{2G}\sqrt{\frac{SG}{\pi}}-\frac{P\sqrt{\frac{SG}{\pi}}}{\sqrt{\beta}G}\kappa\arctan\left(\frac{2P\pi\sqrt{\beta}}{SG}\right)\nonumber \\
 & +\frac{S\sqrt{\frac{SG}{\pi}}}{12\pi\beta}\kappa\ln\left(1+\frac{4P^{2}\beta\pi^{2}}{G^{2}S^{2}}\right)\nonumber \\
 & +\frac{2P^{2}\kappa}{3\sqrt{P\sqrt{\beta}}G}\left[\arctan\left(1-\frac{\sqrt{\frac{SG}{\pi}}}{\sqrt{P\sqrt{\beta}}}\right)-\arctan\left(1+\frac{\sqrt{\frac{SG}{\pi}}}{\sqrt{P\sqrt{\beta}}}\right)\right]\nonumber \\
 & +\frac{P^{2}\kappa}{3\sqrt{P\sqrt{\beta}}G}\ln\left(\frac{2P\sqrt{\beta}+2\sqrt{P\sqrt{\beta}}\sqrt{\frac{SG}{\pi}}+\frac{SG}{\pi}}{2P\sqrt{\beta}-2\sqrt{P\sqrt{\beta}}\sqrt{\frac{SG}{\pi}}+\frac{SG}{\pi}}\right).\label{m_of_bh_interms_of_entropy}
\end{align}
In that case, we redefine the variables of the mass function (\ref{m_of_bh_interms_of_entropy})
as follow: 
\begin{align}
S\rightarrow & \lambda^{i}S,\;P\rightarrow\lambda^{j}P,\;\beta\rightarrow\lambda^{k}\beta,\;\Lambda\rightarrow\lambda^{n}\Lambda,\label{redefined_parameters}
\end{align}
and the integer powers of the redefined parameters take the following
equalities to satisfy the Euler's homogeneous function theorem 
\begin{align}
i=k=2l,\; & j=l,\;n=-2l.\label{power_of_parameters}
\end{align}
With these choice of indices the Smarr's formula becomes 
\begin{align}
M\left(S,P,\beta,\Lambda\right)= & 2S\left(\frac{\partial M}{\partial S}\right)+P\left(\frac{\partial M}{\partial P}\right)+2\beta\left(\frac{\partial M}{\partial\beta}\right)-2\Lambda\left(\frac{\partial M}{\partial\Lambda}\right).\label{Smarr_formula}
\end{align}
The derivative of the black hole mass (\ref{m_of_bh_interms_of_entropy})
with respect to entropy corresponds to the Hawking temperature, 
\begin{equation}
T_{H}=\frac{f^{\prime}\left(r_{h}\right)}{4\pi}=\frac{\partial M}{\partial S},\label{Hawking_temperature}
\end{equation}
where 
\begin{align}
T_{H}=\frac{\partial M}{\partial S}= & \frac{1}{4\sqrt{\pi GS}}+\frac{\sqrt{GS}\Lambda}{4\pi\sqrt{\pi}}-\frac{P\kappa}{2\sqrt{GS\beta\pi}}\arctan\left[\frac{2P\pi\sqrt{\beta}}{GS}\right]\nonumber \\
 & +\frac{\kappa\sqrt{GS}}{8\pi\sqrt{\pi}\beta}\ln\left[1+\frac{4P^{2}\pi^{2}\beta}{G^{2}S^{2}}\right].\label{Der_M_S}
\end{align}
The second term in (\ref{Smarr_formula}) refers to the magnetic potential
$\phi_{P}$ on the horizon

\begin{align}
\phi_{P}=\frac{\partial M}{\partial P}= & \frac{\pi\kappa\sqrt{P}}{\beta^{\frac{1}{4}}G}+\frac{\sqrt{P}\kappa}{G\beta^{1/4}}\left(\arctan\left[1-\frac{\sqrt{\frac{GS}{\pi}}}{\sqrt{P\sqrt{\beta}}}\right]-\arctan\left[1+\frac{\sqrt{\frac{GS}{\pi}}}{\sqrt{P\sqrt{\beta}}}\right]\right)\nonumber \\
 & +\frac{\sqrt{P}\kappa}{2G\beta^{1/4}}\ln\left[\frac{2P\sqrt{\beta}+2\sqrt{P\sqrt{\beta}}\sqrt{\frac{SG}{\pi}}+\frac{SG}{\pi}}{2P\sqrt{\beta}-2\sqrt{P\sqrt{\beta}}\sqrt{\frac{SG}{\pi}}+\frac{SG}{\pi}}\right]-\sqrt{\frac{GS}{\pi}}\frac{\arctan\left[\frac{2\pi P\sqrt{\beta}}{SG}\right]}{\sqrt{\beta}G}.\label{magnetic_potential}
\end{align}
The third term in (\ref{Smarr_formula}) is

\begin{align}
K_{\beta}=\frac{\partial M}{\partial\beta}= & -\frac{\pi\kappa P^{\frac{3}{2}}}{6\beta^{\frac{5}{4}}G}-\frac{P^{3/2}\kappa\left(\arctan\left[1-\frac{\sqrt{\frac{GS}{\pi}}}{\sqrt{P\sqrt{\beta}}}\right]-\arctan\left[1+\frac{\sqrt{\frac{GS}{\pi}}}{\sqrt{P\sqrt{\beta}}}\right]\right)}{6\beta^{\frac{5}{4}}G}\nonumber \\
 & -\frac{P^{3/2}\kappa\ln\left[\frac{2P\sqrt{\beta}+2\sqrt{P\sqrt{\beta}}\sqrt{\frac{SG}{\pi}}+\frac{SG}{\pi}}{2P\sqrt{\beta}-2\sqrt{P\sqrt{\beta}}\sqrt{\frac{SG}{\pi}}+\frac{SG}{\pi}}\right]}{12\beta^{\frac{5}{4}}G}+\frac{P\sqrt{\frac{GS}{\pi}}\kappa\arctan\left[\frac{2\pi P\sqrt{\beta}}{GS}\right]r_{h}}{2\beta^{\frac{3}{2}}G}-\frac{S\sqrt{\frac{GS}{\pi}}\kappa\ln\left[1+\frac{4P^{2}\beta\pi^{2}}{G^{2}S^{2}}\right]}{12\pi\beta^{2}}\label{b_potential}
\end{align}
which is defined as the $\beta$-potential $K_{\beta}$.

The derivative of the black hole mass with respect to $\Lambda$ is
\begin{align}
V=\frac{\partial M}{\partial\Lambda}= & \frac{S\sqrt{\frac{GS}{\pi}}}{6\pi},\label{Der_M_L}
\end{align}
which is nothing but the thermodynamic volume $V$ \cite{Kastor}.

Then the Smarr formula becomes 
\begin{align}
M= & 2ST_{H}+P\phi_{P}+2\beta K+2pV,\label{Smarr_formula_1}
\end{align}
which relates the black hole mass with the entropy, thermodynamic
potentials and the cosmological constant. Note that, in extended thermodynamic
phase space we denoted the cosmological pressure $p$ for $-\varLambda.$
The equation (\ref{Smarr_formula_1}) shows that the first law of
thermodynamics is satisfied for the magnetic black hole solution.
In the next part the thermal stability of the magnetic blackhole solution
is considered.

\subsection{Thermal Stability of the Magnetic Black Hole Solution}

In this part we investigate the thermal stability of the black hole
solution (\ref{exact_solution_magnetic_part}) without the cosmological
constant by analyzing the behavior of the thermal heat capacity when
the magnetic charge is constant while the radius of the event horizon
is changing,

\begin{align}
C_{P}= & T_{H}\left(\frac{\partial S}{\partial T_{H}}\right)_{P},\label{Heat_capacity}
\end{align}
which becomes after defining unitless variables
\begin{align}
\widetilde{C}_{P}= & \frac{\left(2\alpha-2\arctan\left[\frac{1}{x_{h}^{2}}\right]+x_{h}^{2}\ln\left[1+\frac{1}{x_{h}^{4}}\right]\right)x_{h}^{2}}{-2\alpha+2\arctan\left[\frac{1}{x_{h}^{2}}\right]+x_{h}^{2}\ln\left[1+\frac{1}{x_{h}^{4}}\right]},\label{Heat_capacity_unitless}
\end{align}
and Hawking temperature 
\begin{align}
\widetilde{T}_{H}= & \frac{1}{x_{h}}-\frac{1}{\alpha x_{h}}\arctan\left[\frac{1}{x_{h}^{2}}\right]+\frac{x_{h}}{2\alpha}\ln\left[1+\frac{1}{x_{h}^{4}}\right],\label{Hawking_temp_unitless}
\end{align}
where we have defined $r_{0}^{2}\equiv2P\sqrt{\beta}$, the unitless
variables $x_{h}\equiv\frac{r_{h}}{r_{0}}$, $\widetilde{T}_{H}=r_{0}^{4}\pi T_{H},$
$\widetilde{C}_{P}=\frac{G}{2\pi r_{0}^{2}}C_{P}$ and $\alpha=\frac{\sqrt{\beta}}{2P\kappa}$.
The graphs of (\ref{Hawking_temp_unitless}) and (\ref{Heat_capacity_unitless})
are given in Fig.1 and Fig.2 respectively.

\begin{figure}[H]
\caption{\protect\includegraphics[scale=0.6]{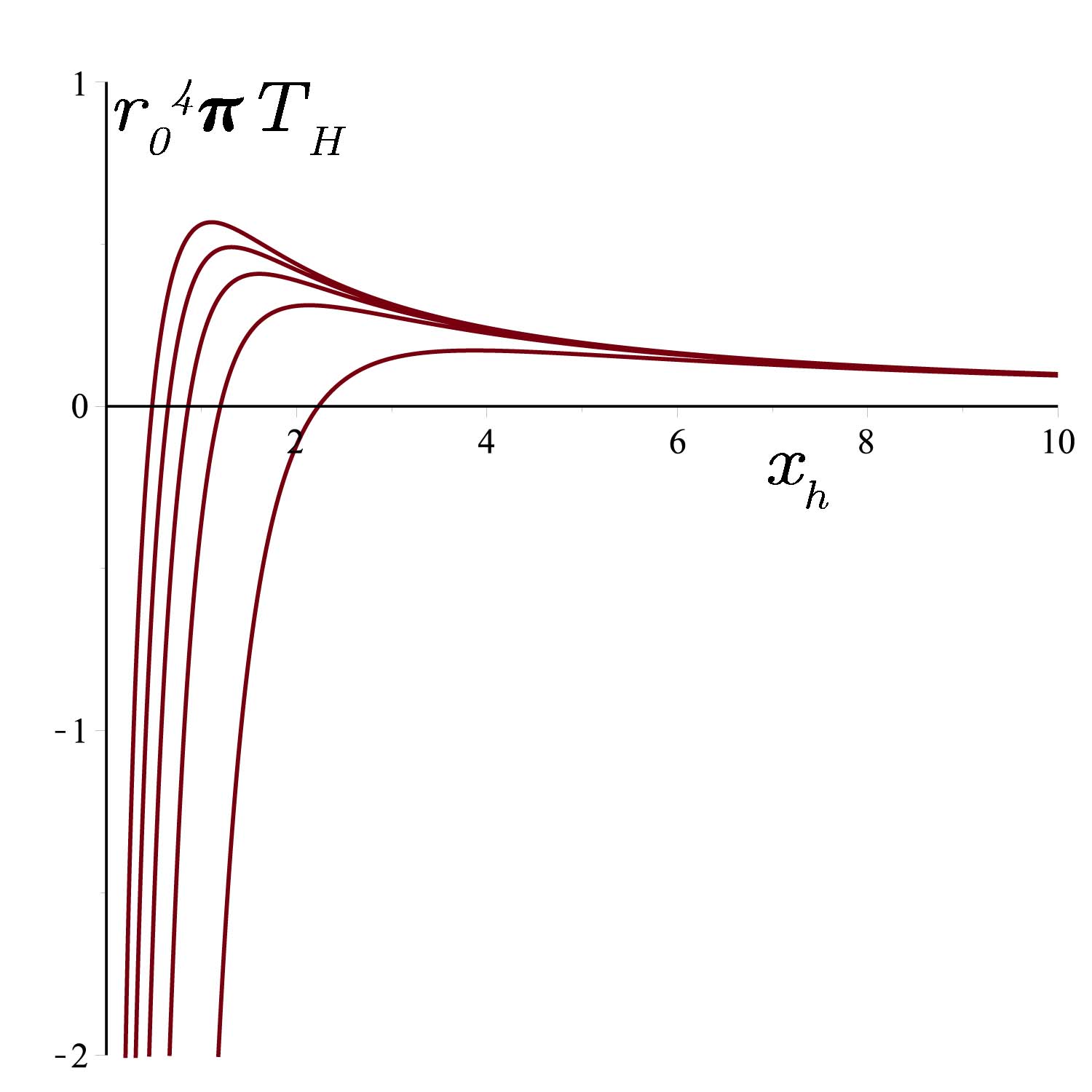}}
The Hawking temperature is plotted with respect to $x_{h}$ for five
different values of $\alpha$ between $0.1$ and $1$ with equal steps.
In all plots we have a maximum value for the Hawking temperature where
the first derivative is zero. 
\end{figure}
The Hawking temperature has only one transition point which is type-1.
When $x_{h}$ is smaller than the transition point the Hawking temperature
is negative and when it is greater than the transition point the Hawking
temperature is positive and it tends to zero for $x_{h}\rightarrow\infty$.

\begin{figure}[H]
\caption{\protect\includegraphics[scale=0.6]{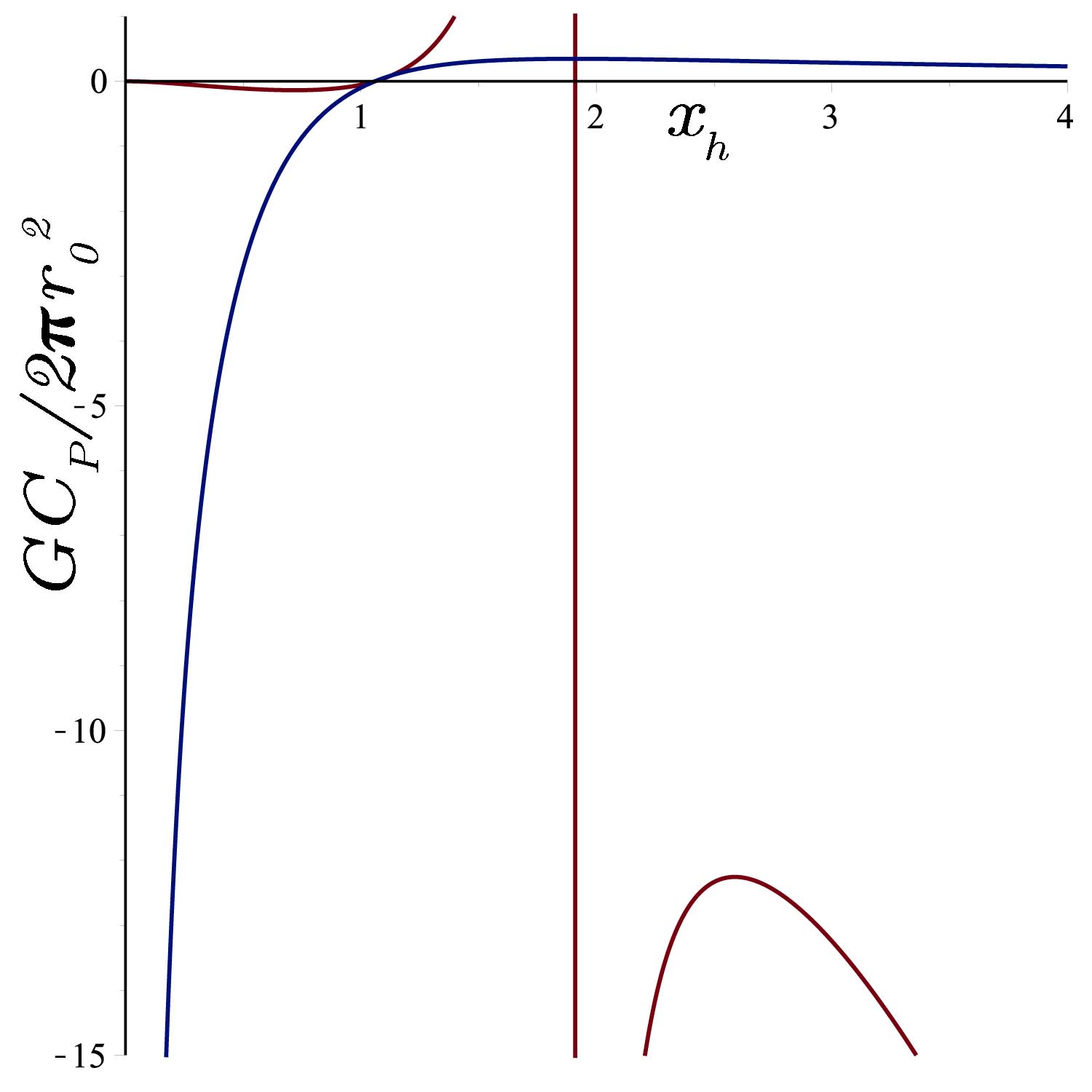}}

Both the Hawking temperature and heat capacity is plotted for $\alpha=0.5$
in this figure. The heat capacity has two transition points one of
which coincides with the transition point of Hawking temperature.
The second transition point of the heat capacity is located at a point
where the Hawking temperature has its maximum value. 
\end{figure}

The positivity of heat capacity and Hawking temperature are sign to
a physical and stable black hole solutions. The heat capacity has
type-1 transition, at which $C_{P}=0$, and type-2 transition, where
$C_{P}\rightarrow\infty$, points (the maximum point of Hawking temperature).
It is negative when $x_{h}$ is between zero and type-1 transition
point and greater than type-2 transition point. It is positive between
the transition points, $C_{P}>0$ which is the necessary condition
to have thermal stability. As a result the black hole is thermally
stable between the transition points where both the heat capacity
and Hawking temperature takes positive values.

\section{Conclusion:}

We have found RN type electric and magnetic black hole solutions in
``Double-Logarithmic nonlinear electrodynamics'' theory which is
given by the Lagrangian (\ref{Lagrangian}). The pure magnetic black
hole solution is given in terms of elementary functions and becomes
singular with a global monopole type constant term which affect the
lensing effect by giving rise to an angle of deficiency. In spite
of that, the pure electric black hole solution can be expressed as
an integral equation. Due to the nice property of magnetic black hole
solution we have analyzed its thermodynamic structure. We have derived
the black hole first law and the Smarr's formula. Also, the stability
is investigated by use of the heat capacity and Hawking temperature.
Both heat capacity and Hawking temperature becomes positive between
type-1 and type-2 transition points of heat capacity where the theory
turns out to be stable and admits stable black holes.

\end{document}